\documentclass[secnumarabic,aps,prb,reprint,groupedaddress]{revtex4-2}

\usepackage[english]{babel}
\usepackage{graphicx}
\usepackage{color}
\usepackage{amsmath}
\usepackage{tabularx}
\usepackage[dvipsnames]{xcolor}

\begin{document}
	
\newcommand{\unit}[1]{\:\mathrm{#1}}            
\newcommand{\To}{\mathrm{T_0}}
\newcommand{\Tp}{\mathrm{T_+}}
\newcommand{\Tm}{\mathrm{T_-}}
\newcommand{\EST}{E_{\mathrm{ST}}}
\newcommand{\Rp}{\mathrm{R_{+}}}
\newcommand{\Rm}{\mathrm{R_{-}}}
\newcommand{\Rpp}{\mathrm{R_{++}}}
\newcommand{\Rmm}{\mathrm{R_{--}}}
\newcommand{\ddensity}[2]{\rho_{#1\,#2,#1\,#2}} 
\newcommand{\ket}[1]{\left| #1 \right>} 
\newcommand{\bra}[1]{\left< #1 \right|} 
	
\title{Effects of the dynamical magnetization state on spin transfer}
\author{Neil Tramsen}
\thanks{N.T. and A.M. contributed equally to this work.}
\author{Alexander Mitrofanov}
\thanks{N.T. and A.M. contributed equally to this work.}
\author{Sergei Urazhdin}
\affiliation{Department of Physics, Emory University, Atlanta, GA, USA}

\begin{abstract}
We utilize simulations of electron scattering by a chain of dynamical quantum spins, to analyze the interplay between the spin transfer effect and the magnetization dynamics. We show that the complex interactions between the spin-polarized electrons and the dynamical states of the local spins can be decomposed into separate processes involving electron reflection and transmission, as well as absorption and emission of magnons - the quanta of magnetization dynamics. Analysis shows that these processes are substantially constrained by the energy and momentum conversation laws, resulting in a significant dependence of spin transfer on the electron's energy and the dynamical state of the local spins. Our results suggest that exquisite control of spin transfer efficiency and of the resulting dynamical magnetization states may be achievable by tailoring the spectral characteristics of the conduction electrons and of the magnetic systems.

\end{abstract}

\maketitle

\section{I. Introduction}
Spin transfer effect (ST) - the transfer of spin angular momentum from spin-polarized conduction electrons to magnetic systems~\cite{Slonczewski1996,Berger1996,Zhang2002,Ralph20081190} - is one of the most extensively studied effects in modern nanomagnetism, thanks to the unique fundamental insights it provides into electron spin physics, a plethora of related magnetoelectronic and dynamical effects, as well as viable applications in information technology~\cite{Kent2015, Divinskiy2016, Locatelli2013,Kim2012,7505988}. ST can result in magnetization reversal~\cite{PhysRevLett.84.3149,Mangin2006}, precession~\cite{Kiselev2003,PhysRevLett.92.027201,Demidov2012} and other dynamical effects~\cite{Demidov2010,Demidov2011,Madami2011}. Magnetic switching driven by ST is finding applications in memories and biomimetics, while its ability to generate magnetization dynamics provides unique opportunities for magnonics - the information and telecommunication technology utilizing the quanta of magnetization dynamics (magnons) as information carriers~\cite{Chumak2015,Grundler2016}.

ST is a consequence of spin angular momentum conservation in the process of scattering of spin-polarized conduction electrons by magnetic systems. This process has been extensively analyzed in the semiclassical approximation for the magnetic systems, in which the magnetic order is approximated as a continuous classical vector field with fixed magnitude, or as an array of localized classical magnetic moments. The latter approximation is commonly utilized  in micromagnetic simulations.  The magnetization dynamics of ferromagnets is usually described by the semiclassical Landau-Lifshitz-Gilbert (LLG) equation with an additional Slonczewski's term arising from ST~\cite{Slonczewski1996}.  The possibility to include ST in the LLG equation, instead of jointly solving the dynamical equations for the conduction electrons and the magnetization coupled by the exchange interaction, requires an adiabatic approximation, i.e., it is assumed that the relevant magnetization dynamics are significantly slower than the spin dynamics of conduction electrons involved in ST. 

We now discuss recent developments in the studies of ST, relevant to the present work, that transcend these approximations. In the semiclassical approximation, the magnitude of magnetization in ferromagnets is fixed, so ST is forbidden by angular momentum conservation if the electron is spin-polarized collinearly with the magnetic order. However, recent experimental measurements~\cite{PhysRevLett.119.257201,Kim2019} and theoretical studies~\cite{PhysRevB.69.134430,PhysRevB.99.094431,PhysRevB.99.024434,petrovic2020, petrovic2020quantum} revealed a contribution to ST, termed the quantum ST, which persists in the collinear geometry. In ferromagnets, this effect becomes noticeable only at cryogenic temperatures. However, it may be dominant in antiferromagnets, and is the only expected contribution to ST in spin liquids, whose magnetic state cannot be described semiclassically~\cite{mitrofanov2020nonclassical,petrovic2020quantum}. 

Theoretical studies have shown that ST leads to quantum entanglement between conduction electrons and magnetization~\cite{PhysRevB.99.094431,petrovic2020}, and can also mediate entanglement within the magnetic system~\cite{petrovic2020quantum}, with possible applications in quantum information technologies~\cite{Tejada_2001, Bou_Comas_2019}. Furthermore, it was shown that the conservation of the total energy and linear momentum of the quasiparticles involved in ST, the electrons and the magnons, can impose 
substantial constraints on the magnetization dynamics induced by ST, and on scattering of electrons by magnetic systems~\cite{alex2020energy}. While the roles of energy and linear momentum in ST were analyzed already in the semiclassical models~\cite{PhysRevB.88.144413, PhysRevB.98.224401, PhysRevLett.92.086601}, the constraints imposed by the magnon energy and momentum, in the de Broglie sense, were not captured by these models. 

In this work, we utilize quantum simulations of electron scattering by a quantum spin chain initially populated with one magnon, to analyze the effects of magnetization dynamics on ST. In principle, such effects can be introduced already in the semiclassical approach, by jointly solving the coupled dynamical equations for the conduction electrons and the magnetization~\cite{PhysRevB.101.214412}. However, our quantum simulations show that the conservation of the total energy and linear momentum (in the de Broglie sense) of the quasiparticles involved in ST, the electrons and the magnons,  plays a central role in the studied phenomena. Thus, our results provide further evidence for the significance of non-classical aspects of ST, and suggest a new route for controlling its efficiency. 

The rest of this paper is organized as follows. In Section II, we describe the model and provide the computational details. In Section III, we classify and analyze different scattering processes involved in ST, for a spin chain populated with one magnon. In Section IV, we consider ST for the electron spin-polarized orthogonally to the equilibrium magnetization, and show that this case, common in the ST studies, includes all the contributions analyzed in Section III. We summarize our observations in Section V.

\section{II. Model and simulation details}

We consider an electron initially propagating in a non-magnetic medium, and subsequently scattered by a ferromagnet (FM) modeled as a chain of $n=10$ localized spins-1/2. In the tight-binding approximation, this system can be described by the Hamiltonian~\cite{PhysRevB.99.094431,alex2020energy}
\begin{equation}\label{eq:hamiltonian}
\begin{split}
\hat{H}=-\sum_{i}b|i\rangle\langle i+1|\\
-\sum_{j}(J\hat{\mathbf{S}}_j\cdot\hat{\mathbf{S}}_{j+1}+J_{sd}|j\rangle\langle j|\otimes\hat{\mathbf{S}}_j\cdot\hat{\mathbf{s}}),
\end{split}
\end{equation}
where the first term on the right describes hopping of the itinerant electron, the second - its exchange interaction with the local spins, and the last term - the exchange between the local spins. Indices $i=1...180$ in Eq.~(\ref{eq:hamiltonian}) enumerate the tight-binding sites of the entire considered system, including the FM and the non-magnetic medium surrounding it, while indices $j=70-80$ enumerate the sites occupied by the localized spin-1/2 chain representing the FM.  $\hat{\mathbf{S}}_j$, $\hat{\mathbf{s}}$ are the spin operators of the electron and of the local spins, $b$ is the electron hopping parameter, $J$ is the exchange stiffness of the local spins, and $J_{sd}$ is their exchange interaction with the electron. The calculations below use $b=1$~eV, $J=J_{sd}=0.1$~eV, unless specified otherwise.

We use periodic boundary conditions for both the electron and the spin chain, to avoid artifacts associated with reflections at the boundaries. Spin-orbit interactions are neglected in our model. Nevertheless, we expect our results to be broadly relevant to spin-orbit torques, thanks to the general validity of conservation laws governing electron-magnon scattering.  For typical experimental magnetic fields, the effects of the Zeeman interaction are negligible on the considered time scales, aside from defining the quantization axis for the local spin dynamics. In the following, we assume that the local spins are aligned with the z-axis in their ground state, with $\left<S_z\right>=5\hbar$. The same value $a$ of the lattice constant, which defines the possible values of quasiparticle wavevectors, is used throughout the entire system. 

To analyze ST, the system is initialized with the electron forming a Gaussian wavepacket centered around the wavevector $k_e^{(i)}$ and polarized in the $+z$, $-z$, or $x$  direction. The local spins are populated with one magnon with wavevector $k_m^{(i)}$. In this state, $\left<S_z\right>=4\hbar$. The system is then evolved using the Schrodinger equation with the Hamiltonian Eq.~(\ref{eq:hamiltonian}). We choose the width of the electron wavepacket so that it remains well-defined throughout the scattering process, allowing us to clearly identify the time intervals corresponding to its propagation in the non-magnetic medium before and after ST, and enabling us to unambiguously determine the associated changes of physical quantities.

To analyze the evolution of the two subsystems, we introduce the density matrices $\hat{\rho}_e=\mathrm{Tr_m}\hat{\rho}$ and $\hat{\rho}_m=\mathrm{Tr_e}\hat{\rho}$ for the electron and the local spins, respectively, by tracing out the full density matrix $\hat{\rho}$ with respect to the other subsystem~\cite{PhysRevB.99.094431}. The expectation value of a physical quantity $\hat{A}$ associated with the electron is $\left<\hat{A}\right>=\mathrm{Tr}(\hat{A}\hat{\rho}_e)$, while the probability of its value $a$ is $P_a=\left<\psi_a|\hat{\rho}_e|\psi_a\right>$, where $\psi_a$ is the corresponding eigenstate. Similar relations are used to analyze the observables associated with the local spins. For instance, the expectation values of different contributions to the system’s energy are obtained by using the corresponding terms in the Hamiltonian Eq.~(\ref{eq:hamiltonian}) as $\hat{A}$. The distribution of electron momentum $p_e$ is obtained by projecting onto the plane-wave eigenstates $|\psi_k\rangle\sim~e^{ik_ex}$. Here, $k_e=p_e/\hbar$ is the wavevector describing the corresponding Fourier component of the electron wave. For brevity, we interchangeably use the terms ''wavevector" (or ''wavenumber") and ''momentum" (in the de Broglie sense), since the two quantities are simply related by the Planck's constant $\hbar$. For magnons, we utilize the Bethe ansatz to classify the eigenstates, as described below, and project $\hat{\rho}_m$ onto these states to determine the distribution of magnon populations and their energies/momenta.

\begin{figure}
	\includegraphics[width=\columnwidth]{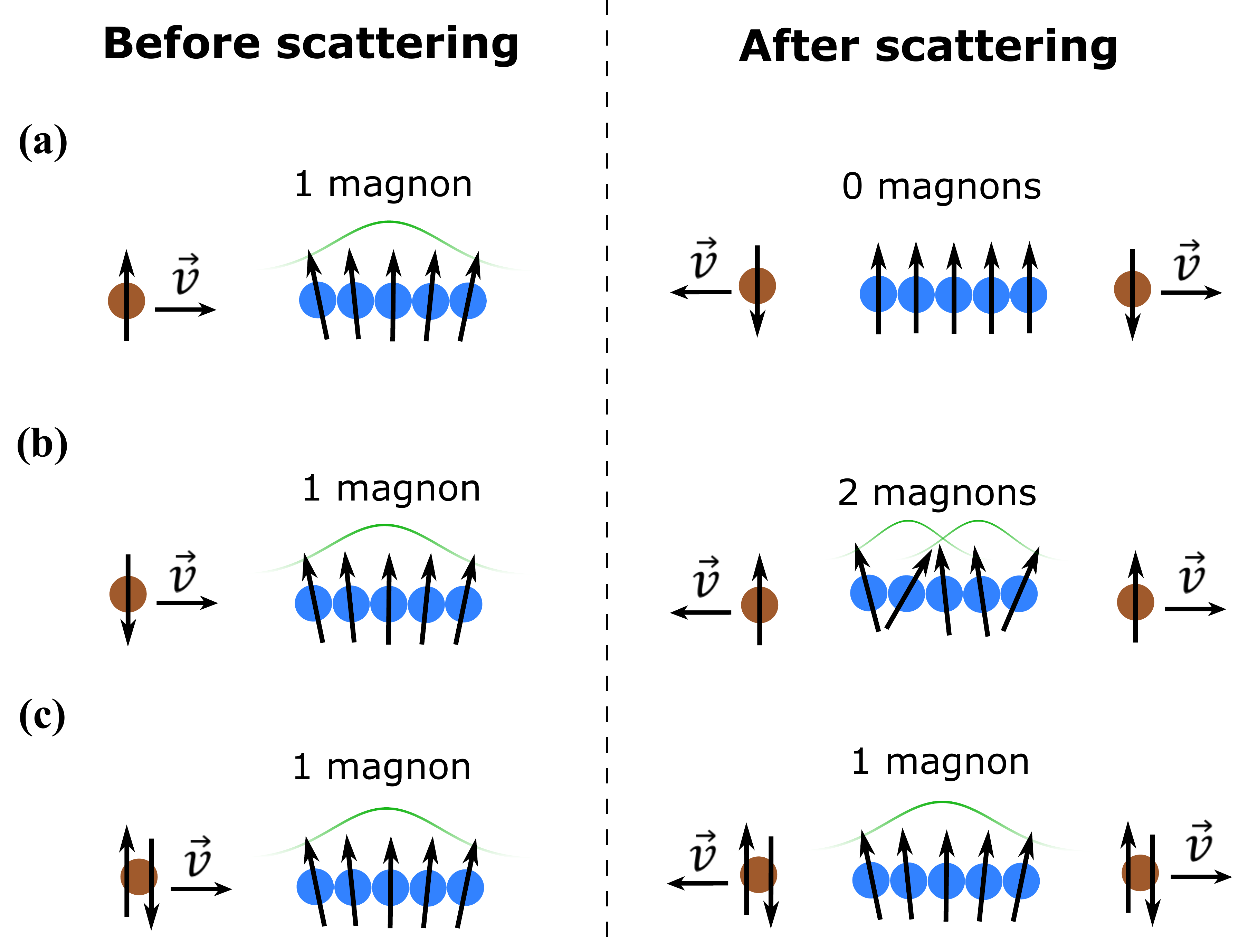}	
	\caption{\label{fig:main_schematic} (Color online) Processes involved in ST, for the local spins populated with one magnon. (a) For a spin-up incident electron, the magnon can be absorbed, flipping electron spin to down. (b) For a spin-down incident electron, an additional magnon can be emitted, flipping electron spin to up. (c) The electron can be scattered without spin-flipping or changing the number of magnons in the system, but nevertheless exchange energy and momentum with the existing magnon. In all cases, the electron can be either reflected or transmitted.}
\end{figure}

\section{III. Classification and analysis of the scattering processes involved in ST}

In this Section, we identify and characterize several scattering processes involved in ST, for a spin chain initially populated with one magnon, and demonstrate that these processes lead to distinct outcomes. For clarity, Fig.~\ref{fig:main_schematic} illustrates these processes  separately for the spin-up and spin-down polarization (i.e., polarization in the $+z$ and $-z$  directions) of the incident electron. In the next Section, we show that scattering of the incident electron polarized in the x-direction can be interpreted as a superposition of all these processes, demonstrating their relevance for the generic ST geometries involving spin currents non-collinear with the magnetization. 

The incident electron can be either transmitted or reflected, and its polarization can either change or remain the same. Spin-up is parallel to the local spins in their ground state. Since each magnon carries spin 1, angular momentum conservation requires that electron scattering either results in the absorption of the initially present magnon (panel a), or one magnon remains in the system (panel c). In the former case, the electron must spin-flip to spin-down, while in the latter its spin must remain the same. We note that even if the magnon population does not change, energy and linear momentum can be exchanged between the electron and the magnon as a result of scattering.

By the same spin angular momentum conservation argument for the incident spin-down electron, ST can result in the generation of a second magnon accompanied by electron spin-flipping into the spin-up state  [Fig.~\ref{fig:main_schematic}(b)]. The electron can be also scattered without ST, but nevertheless exchange energy and momentum with the magnon, similarly to the spin-up electron.

\begin{figure}
	\includegraphics[width=\columnwidth]{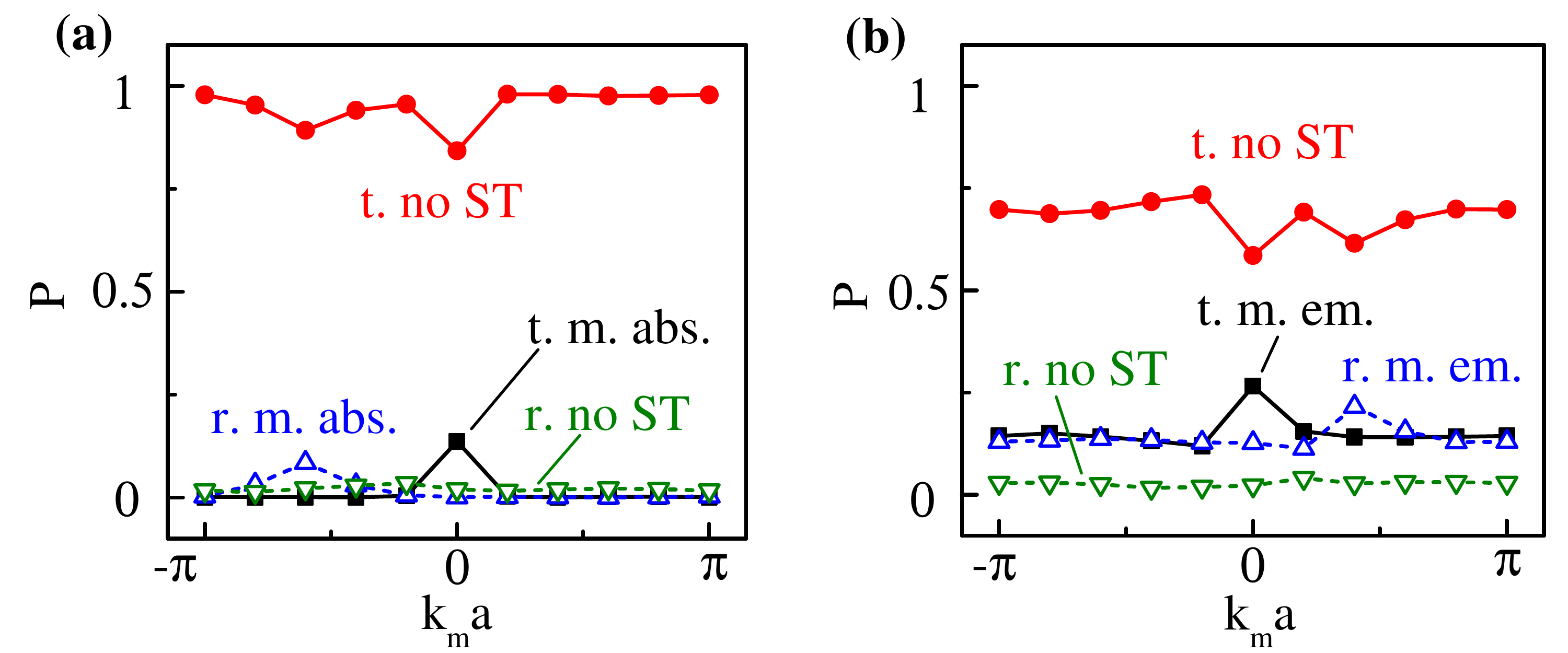}	
	\caption{\label{fig:schematic_probabilities} (Color online) Probabilities of different scattering outcomes, as classified in Fig.~\ref{fig:main_schematic}, for spin-up (a) and spin-down (b) incident electron vs $k_m^{(i)}a$, for $k_e^{(i)}a=0.6$. In the abbreviated labels, electron transmission and reflection is denoted as ``t." and ``r.", respectively, while ``m. abs.", ``m. em.", and ``no ST" denote the cases of magnon absorption, magnon emission, and the absence of ST, respectively.}
\end{figure}

 We confirmed the scenarios identified in Fig.~\ref{fig:main_schematic} by projecting the results of the simulations onto the eigenstates of the system, as described in Section II. The dependencies of the probabilities of different scattering outcomes on the initial magnon wavevector are shown for spin-up and spin-down incident electrons in Figs.~\ref{fig:schematic_probabilities} (a) and (b), respectively.
 
We note several important features of scattering. First, different scattering processes contribute differently to ST. For instance, absorption by a spin-up electron of a magnon with wavenumber $k_m^{(i)}=0$  is accompanied by electron transmission (labeled ``t. m. abs."). Second, the probabilities of different processes are strongly dependent on the magnitude of the magnon wavevector. For instance, for the spin-up polarization of the incident electron, the probability of electron transmission without ST (labeled ``t. no ST" in Fig.~\ref{fig:schematic_probabilities}(a)) is close to $1$ for most values of $k_m>0$, except for a pronounced dip around $k_m=0$. Meanwhile, the probability of reflection without ST, labeled ``r. no ST", remains negligible at all $k_m^{(i)}$. As a consequence, the total probability of scattering without ST, and conversely the magnitude of ST, is dependent on $k_m^{(i)}$. A clear asymmetry of these results with respect to the sign of $k_m^{(i)}$ indicates that both the magnitude and the direction of the magnon momentum play important roles in the scattering processes.

The dominance of transmission without ST among the scattering processes for spin-up electrons can be qualitatively understood as a consequence of weak effects of the $s$-$d$ interaction between the conduction electron's spin and the local spins that are almost parallel to each other. These effects are significantly stronger for the spin-down incident electron. In particular, while electron transmission without ST is still dominant, the probability of ST accompanied by both electron transmission and reflection is significantly larger [Fig.~\ref{fig:schematic_probabilities}(b)]. We also note that the dependencies on $k_m^{(i)}$ for the spin-down electron are substantially different from those for the spin-up electron, suggesting that ST involves a complex interplay between the spin and the orbital degrees of freedom of the magnetic system and of the electron.

\subsection{A. Classification of the dynamical states of the magnetic system and of the electron}

The results of Fig.~\ref{fig:schematic_probabilities} demonstrate that electron scattering and ST are strongly affected by the dynamical state of the magnetic system. We now quantitatively analyze these effects, and show that they are governed by the conservation of energy and linear momentum, in the de Broglie sense, of quasiparticles involved in ST - the electrons and the magnons~\cite{alex2020energy}.

To determine the characteristics of magnons generated by ST, we classify the eigenstates of the magnetic system using the Bethe ansatz~\cite{bethe1931theorie, karbach1998introduction}. The one-magnon states are plane waves with wave numbers $k_m$:
\begin{equation}\label{eq:1_magnon}
\ket{\psi}=\sum_{x=1}^n\frac{1}{\sqrt{n}}\exp{(ik_max)}\ket{x},
\end{equation} 
where $n$ is the number of the local spins, and $a$ is the lattice constant. Inserting this ansatz into the Hamiltonian, we obtain the dispersion relation:
\begin{equation}\label{eq:disp_law_magnon}
E_m=4J(1-\cos{k_ma})-E_0.
\end{equation} 
where $E_0=-Jn$ is the ground state energy of FM.

The two-magnon states are characterized by two wavenumbers $k_1$, $k_2$ that are either real or complex-conjugates, and satisfy $k_1+k_2\equiv k_{2m}=2\pi/n(\lambda_1+\lambda_2)$, where $\lambda_i$ are integer Bethe numbers~\cite{karbach1998introduction}. The values of $k_1$, $k_2$ are obtained numerically by plugging the Bethe ansatz for the two-magnon states into the Hamiltonian. The eigenenergies have the form
\begin{equation}\label{eq:disp_law_two_magnon}
E_{2m}(k)=4J\sum_{i}(1-\cos{k_{i}a})-E_0,
\end{equation}
where $i=1,2$. For $n=10$ spins in the chain, there are 45 two-magnon states.

For the electron wavepacket centered at the wavevector $k_e$ and localized outside the magnetic system before or after scattering, the dispersion is
\begin{equation}\label{eq:disp_law_electron}
	E_e=2b(1-\cos{k_ea}),
\end{equation} 
as determined from the hopping term in the Hamiltonian. 

We will now utilize this classification of the dynamical states to separately analyze the distinct scattering processes involved in ST.

\subsection{B. Magnon absorption}

\begin{figure}
	\includegraphics[width=\columnwidth]{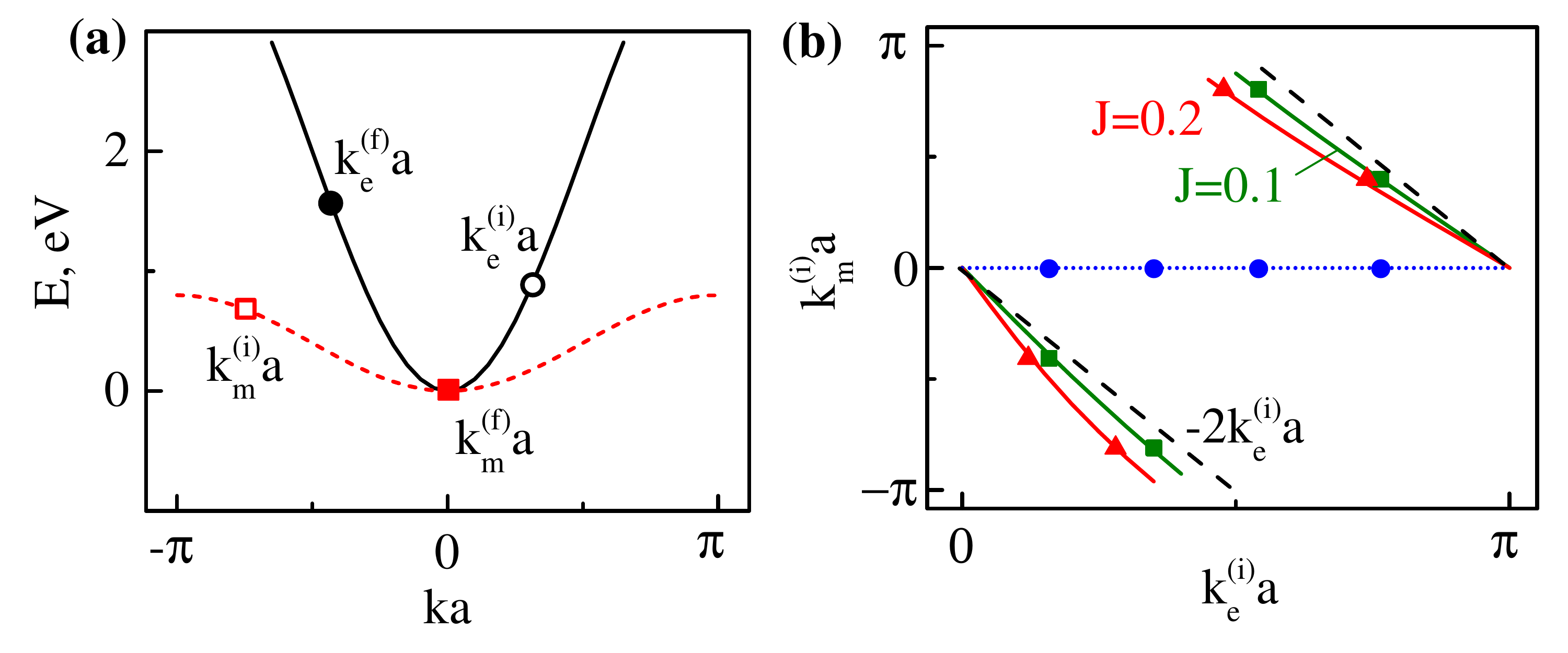}	
	\caption{\label{fig:spin_absorption} (Color online) Analysis of the magnon absorption process. (a) Electron dispersion (solid curve) and magnon dispersion (dashed curve). The states of the subsystems before and after scattering are shown with open and filled symbols, respectively, for $k_m^{(i)}=-2.3/a$, $k_e^{(i)}=1/a$ allowing magnon absorption. (b) Relationship between the electron momentum and the magnon momentum allowing magnon absorption, for the labeled values of $J$.	Curves: numerical solution of Eqs.~(\ref{eq:E_abs}), (\ref{eq:k_abs}), symbols: results of simulations. Dashed lines show $k_m^{(i)}=-2k_e^{(i)}$ and $k_m^{(i)}=-2k_e^{(i)}+2\pi/a$.}
\end{figure}

An incident spin-up electron can absorb a magnon, flipping its spin, and bringing the magnetic system to its ground state [Fig.~\ref{fig:main_schematic}(a)]. As Fig.~\ref{fig:schematic_probabilities}(a) illustrates, for the transmitted electron, the probability of this process is maximized for the initial magnon momentum $k_m^{(i)}=0$. Meanwhile, for the reflected electron, the probability is maximized at large negative $k_m^{(i)}$. These observations can be explained by the constraints imposed by the conservation of energy and linear momentum, in the de Broglie sense, as follows.

Since the Hamiltonian Eq.~(\ref{eq:hamiltonian}) is time-independent, the total energy of the system comprising the electron and the spin chain must be conserved. This conservation law takes the simplest form when the electron is localized in the non-magnetic medium before or after scattering and the two subsystems do not interact, 
\begin{equation}\label{eq:E_abs}
 E_{e}^{(i)}+E_{m}^{(i)}=E_{e}^{(f)},
\end{equation}
where $E_{e}$ ($E_{m}$) is the electron (magnon) energy, in the de Brogile sense, and superscripts $(i)$, $(f)$ denote the characteristics of the quasiparticle before and after scattering, respectively.

For the momentum (or more precisely, quasi-momentum for the discrete lattice), the situation is more complicated, because the spin chain breaks the translation symmetry of the electron's spatial domain, so the momentum needs not be conserved. Nevertheless, the conservation of linear momentum can be interpreted as a consequence of the constructive wave interference among the quasiparticles involved in the scattering process. This condition can be expressed by~\cite{alex2020energy}

\begin{equation}\label{eq:k_abs}
	k_{e}^{(i)}+k_{m}^{(i)}=k_{e}^{(f)}+2\pi l/a,
\end{equation}
where $k_{e}$ ($k_{m}$) is the electron (magnon) wavenumber, and the last term with integer $l$ accounts for the umklapp processes.

We note that interference takes place inside the spin chain, while $k_{e}^{(i)}$ and $k_{e}^{(f)}$ are defined outside the chain. The s-d exchange may be expected to result in the shift of the electron's momentum as it crosses the boundary of the spin chain. It is common to account for such effects by using the approximation of s-d exchange-induced conduction band splitting. However, the s-d exchange term in the Hamiltonian Eq.~(\ref{eq:hamiltonian}) is itself the mechanism of ST discussed in this work, i.e., at small $J_{sd}$ it can be considered as a perturbation for an electron whose dispersion is not modified by s-d exchange. In quantum-mechanical terms, the complexity of this problem stems from the quantum entanglement between the conduction electron and the local spins, such that single-quasiparticle dispersion relations become insufficient to describe the dynamics of the system~\cite{PhysRevB.99.094431,petrovic2020}. However, these effects are small if the electron energy is substantially larger than the s-d exchange energy, and will be neglected here.

 Equations (\ref{eq:E_abs}) and (\ref{eq:k_abs}), with the quasiparticle energies related to their momenta by dispersions Eqs.~(\ref{eq:disp_law_magnon}), (\ref{eq:disp_law_electron}), give two possible values for $k_m^{(i)}$ for a given $k_e^{(i)}$ (or vice versa) that permit magnon absorption.  Since  magnon dispersion is gapless~\cite{comment2}, the magnon with $k_m^{(i)}=0$, $E_m^{(i)}=0$ can be absorbed, while the electron is transmitted without changing its energy or momentum. 
 
 To analyze the second possibility, we consider the electron and the magnon dispersions [Fig.~\ref{fig:spin_absorption}(a)]. Electrons are significantly more dispersive than magnons. If  magnon dispersion were negligible, the electron could be elastically reflected while absorbing a magnon with $k_m^{(i)}=-2k_e^{(i)}+2\pi l/a$, where integer $l$ accounts for the umklapp processes. Non-negligible magnon dispersion results in an increase of the reflected electron's energy, while the absorbed magnon's momentum shifts in the negative direction. Symbols in Fig.~\ref{fig:spin_absorption}(a) show the  momenta and energies of quaisparticles before and after scattering, obtained from the simulations for $k_e^{(i)}=1/a$, $k_m^{(i)}=-2.3/a$ allowing magnon absorption. The relations among the momenta and energies are consistent with our analysis.

Figure~\ref{fig:spin_absorption}(b) shows the dependence of the wavenumber of the absorbed magnon on the wavenumber of the incident electron, both for the transmitted and for the reflected electrons. Calculations based on the conservation laws [curves], are in agreement with the results of quantum simulations [symbols]. The transmitted electron absorbs a zero-momentum magnon, regardless of $k_e^{(i)}$ or the exchange stiffness $J$, as expected from energy and momentum conservation [horizontal  line  in Fig.~\ref{fig:spin_absorption}(b)]. 

For the reflected electron, the dependence $k_m^{(i)}(k_e^{(i)})$ is slightly shifted below the linear form $k_m^{(i)}=-2k_e^{(i)}+2\pi l/a$ expected for negligible magnon dispersion [dashed lines in Fig.~\ref{fig:spin_absorption}(b)], with an abrupt jump close to $k_e^{(i)}\lesssim \pi/2a$ due to umklapp. The downward shift increases with increasing $J$ that controls magnon dispersion, consistent with the discussed mechanisms.
 
The two main takeaways from our analysis of magnon absorption are i) this process is selective with respect to the magnon characteristics. The selectivity is controlled by the electron's energy and momentum, and ii) reflected electrons contribute to  this process differently from the transmitted electrons. 

\subsection{C. Scattering without spin transfer}

Studies of the effects of spin-polarized currents on magnetic systems usually focus on the transfer of spin. From this perspective, no effects on the state of the magnetic system would be expected in the absence of ST. We show below that this is not the case. While the magnon population does not change in the absence of ST,  magnon energy can be modified by electron scattering. The results discussed below were obtained for scattering of the spin-up electron. They were similar for the spin-down incident electron.

In the absence of ST, the energy and momentum conservation relations are 
\begin{equation}\label{eq:Ek_noST}
\begin{split}
E_e^{(i)}+E_m^{(i)}=E_e^{(f)}+E_m^{(f)},\\
k_e^{(i)}+k_m^{(i)}=k_e^{(f)}+k_m^{(f)}+2\pi l/a,
\end{split}
\end{equation}
with integer $l$ accounting for umklapp. For given $k_m^{(i)}$ and $k_e^{(i)}$, Eqs.~(\ref{eq:Ek_noST}) give two solutions for the final state.

\begin{figure}
	\includegraphics[width=\columnwidth]{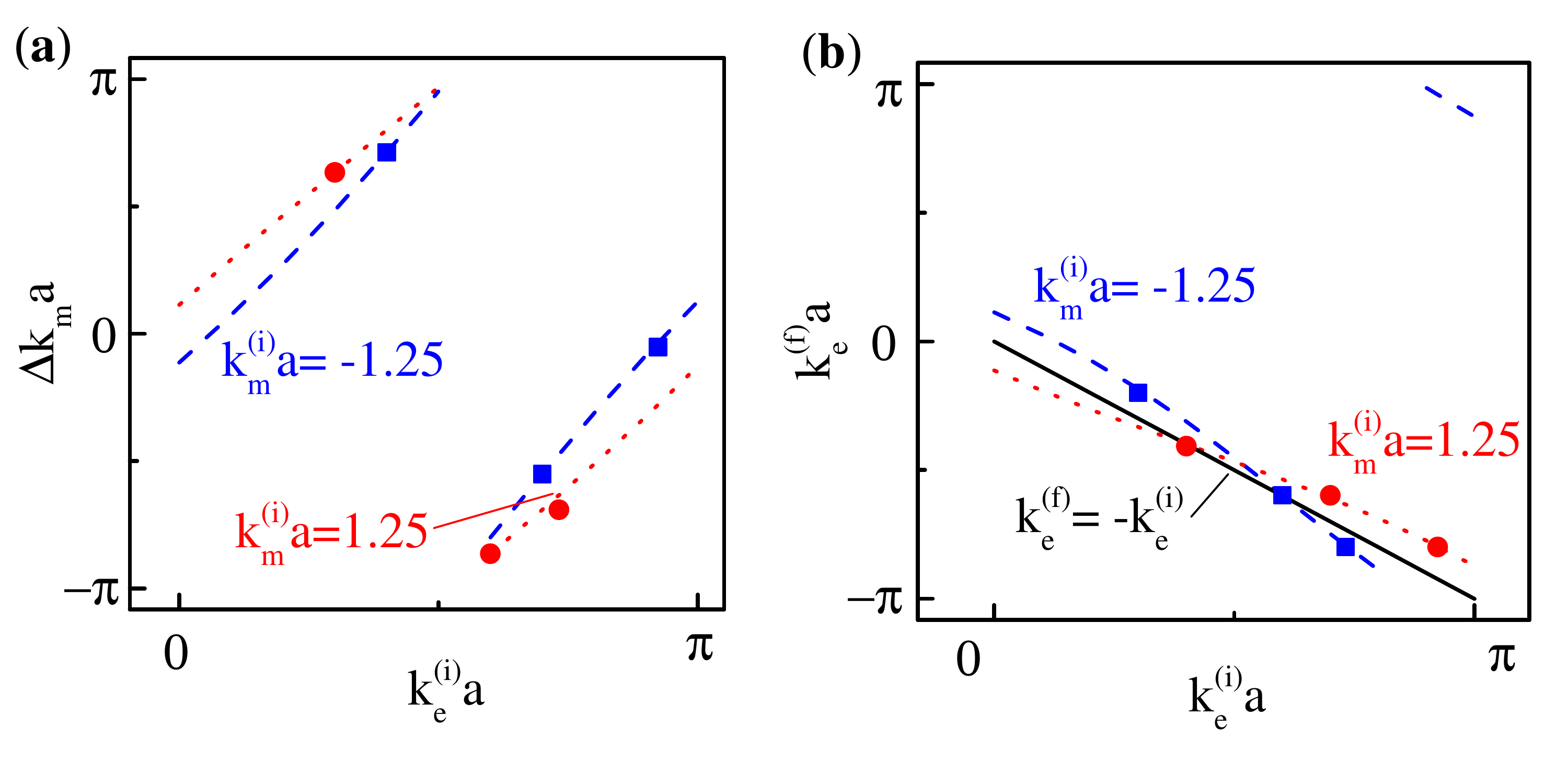}	
	\caption{\label{fig:scattering_without_ST} (Color online) Effects of electron scattering without ST. (a) The change of magnon's wavenumber vs the wavenumber of the incident electron, (b) The wavenumber of the electron after scattering vs its initial wavenumber. Curves are calculations based on the conservation laws Eqs.~(\ref{eq:Ek_noST}), symbols are the results of simulations. Dashed dotted curves are for $k_{m}^{(i)}=-1.25/a$ and $k_{m}^{(i)}=1.25/a$, respectively, the solid line is $k_e^{(f)}=-k_e^{(i)}$. The trivial forward-scattering solution $k_e^{(f)}=k_e^{(i)}$ is not shown.}
\end{figure}

One of the two solutions is trivial - the wavenumbers of both the electron and the magnon remain the same, i.e., it is an elastic forward-scattering process. Naively, one may expect that the second solution must correspond to a similar elastic electron reflection. However, the reversal of the sign of electron's wavevector must be associated with the exchange of momentum, and consequently of energy, between the electron and the magnetic system, resulting in the modification of the magnon properties even without ST.

Figure~\ref{fig:scattering_without_ST}(a) shows the change of the magnon's wavenumber as a function of the incident electron's wavenumber, for two opposite values of the initial magnon wavenumber. These results demonstrate that non-ST electron scattering results in a large magnon drag effect - a shift of the magnon momentum  determined by the initial momentum of the electron. At small $k_e^{(i)}$, the magnon's wavenumber is shifted in the direction of incident electron's momentum. At large $k_e^{(i)}$, the shift switches to the opposite direction, due to the onset of umklapp process at $k_e^{(i)}$ close to $\pi/2a$.

The reciprocal effect on the scattered electron is illustrated in Fig.~\ref{fig:scattering_without_ST}(b), which shows the dependence of the electron's wavenumber after scattering on its initial wavenumber.  At $k_e^{(i)}<\pi/2a$, the wavenumber of the scattered electron is shifted in the direction opposite to the wavenumber of the magnon, relative to the dependence $k_e^{(f)}=-k_e^{(i)}$ expected for the elastic electron back-scattering. At larger $k_e^{(i)}$, the sign of the shift is reversed due to the onset of magnon umklapp.

For most values of $k_e^{(i)}$, the electron is backscattered, $k_e^{(f)}<0$. However, for $k_e^{i}$ close to the center or the boundary of the Brillouin zone and $k_m^{i}<0$, the electron becomes forward-scattered. This effect can be described as suppression of electron backscattering by the magnetic material due to the constraints imposed by the conservation laws. 

For parameters corresponding to the transition between forward- and back-scattering, the velocity of the scattered electron vanishes. 
This outcome may be particularly useful for current-driven phenomena, for two reasons. First, all the initial kinetic energy of the electron is transferred to the magnetic system. Second, the scattered electron remains in the magnetic system for a long period of time, increasing the probability of magnon generation due to the s-d exchange.

\subsection{D. Excitation of two-magnon states}

\begin{figure}
	\includegraphics[width=\columnwidth]{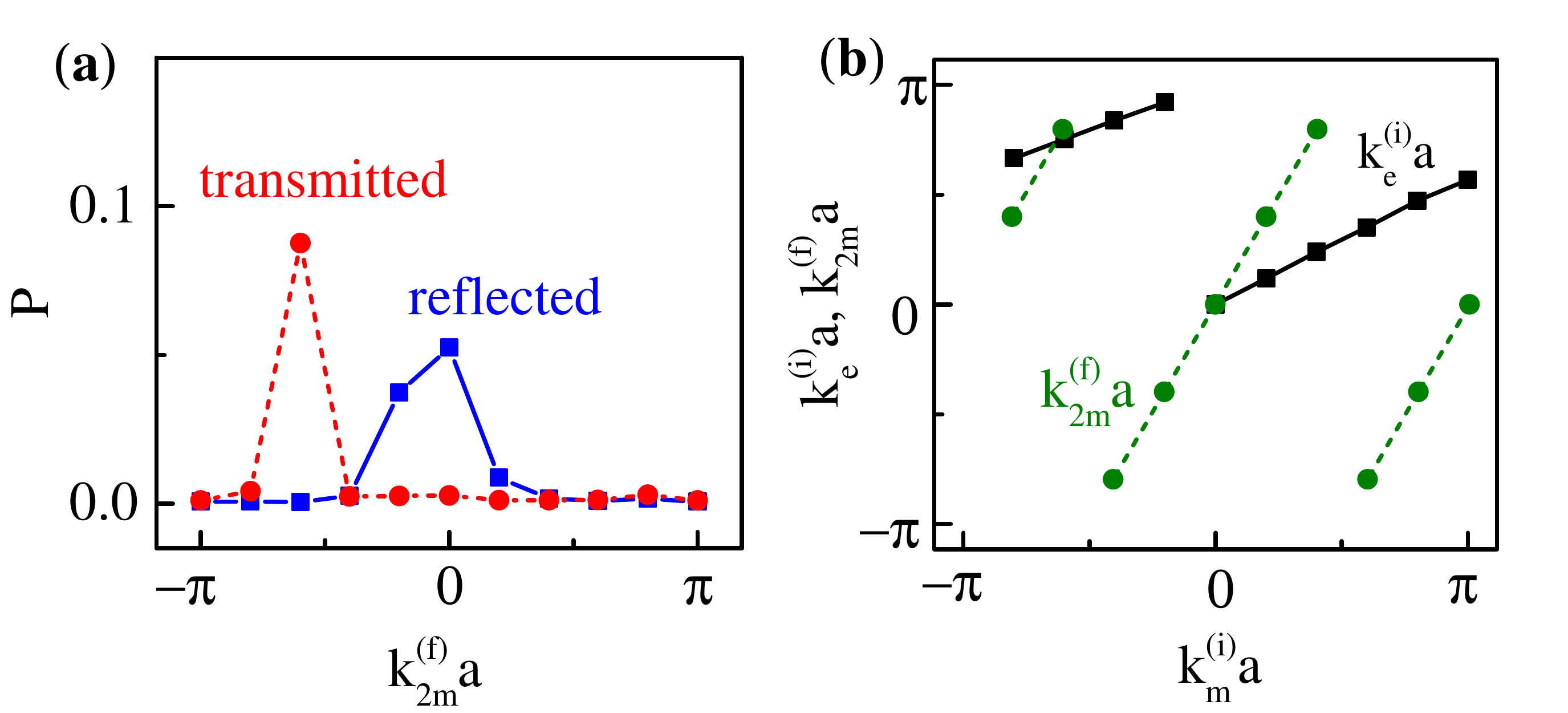}	
	\caption{\label{fig:2magnons} (Color online) Excitation of the two-magnon state by the spin-down electron. (a) Probability distribution of the two-magnon wavenumber $k_{2m}$ for the reflected (squares connected by solid lines) and transmitted electron components (circles connected by dashed lines), at $k_e^{(i)}=1/a$, $k_m^{(i)}=1.88/a$. (b) The resonant value of $k_e^{(i)}$ vs $k_m^{(i)}$, and the resulting values of the two-magnon wavevector $k_{2m}^{(f)}$.}
\end{figure}

An incident electron polarized in the $-z$ direction can excite two-magnon states [Fig.~\ref{fig:main_schematic}(b)]. In this case, one can expect the conservation laws to be less restrictive than in magnon absorption or magnon number-conserving processes, because of the additional degrees of freedom of two magnons in the final state. Indeed, the probability of two-magnon excitation remains finite for all $k_m^{(i)}$ at a given $k_e^{(i)}$, both for the reflected and for the transmitted electron component, as illustrated by the curves labeled  "r. m. em." and "t. m. em." in Fig.~\ref{fig:schematic_probabilities}(b). Nevertheless, these curves exhibit sharp peaks resulting from the conservation laws, as follows.

The energy conservation relation for the two-magnon state excitation is
\begin{equation}\label{eq:E_em}
E_e^{(i)}+E_m^{(i)}=E_e^{(f)}+E_{2m}^{(f)},
\end{equation} 
where $E_{2m}^{(f)}$ is the energy of the two-magnon state given by the Bethe ansatz Eq.~(\ref{eq:disp_law_two_magnon}).The wavefunction of the two-magnon state can be characterized by the two-magnon wavenumber $k_{2m}=k_1+k_2=2\pi/n(\lambda_1+\lambda_2)$, where $n=10$ is the size of the spin chain, and $\lambda_i$ are the integer Bethe numbers~\cite{karbach1998introduction}. The values of $k_1$, $k_2$ can be complex due to the magnon-magnon interaction, so for some two-magnon states they cannot be interpreted as single-magnon wavevectors. The momentum conservation relation is
\begin{equation}\label{eq:k_em}
k_e^{(i)}+k_m^{(i)}=k_e^{(f)}+k_{2m}^{(f)}+2\pi l/a.
\end{equation} 

The conservation relations Eqs.~(\ref{eq:E_em}), (\ref{eq:k_em}) do not prevent excitation of the two-magnon state for any given $k_m^{(i)}$,  $k_e^{(i)}$. For instance, for the transmitted electron, there is always a solution $k_{2m}^{(f)} = k_m^{(i)}$ with $k_1 = 0$ and $k_2 = k_m^{(i)}$, with $E(k^{(f)}_{2m})=E(k_m^{(i)})$. Nevertheless, these relations result in well-defined characteristics of scattered electron and of the excited two-magnon state, as illustrated in Fig.~\ref{fig:2magnons}(a) for the two-magnon momentum, for a generic set of initial conditions. As expected, for the transmitted electron component, the two-magnon momentum is the same as the initial magnon momentum. For the reflected electron, the final-state momentum is determined by the dispersions of the involved quasiparticles.

The probability of two-magnon excitation is strongly dependent on the initial state. For the transmitted electron, it is maximized for $k_m^{(i)}=0$, i.e. when $k_1=k_2=k_m^{(i)}=0$ - the additional excited magnon has the same wavevector as the initial magnon wavevector, which remains unchanged [see the curve labeled ''t. m. em." in Fig.~\ref{fig:schematic_probabilities}(b)]. This result can be interpreted as stimulated magnon emission, i.e., electron spin flip-driven emission of an additional magnon with the same characteristics as the magnon(s) initially in the system. Stimulated magnon emission is the quantum-mechanical picture for the semi-classical spin torque~\cite{Berger1996,PhysRevB.69.134430,PhysRevLett.119.257201}. 

The situation is more complicated for the reflected electron. Figure~\ref{fig:2magnons}(b) shows the dependence of $k_e^{(i)}$ on  $k_m^{(i)}$ maximizing the probability of two-magnon excitation by the reflected electron. The ''resonant" condition is $k_e^{(i)}\approx k_m^{(i)}/2+\pi l/a$, such that the linear momentum conservation relation Eq.~(\ref{eq:k_em}) gives $k_{2m}^{(f)}=2k_m^{(i)}+2\pi l'/a$, where $l'$ is an integer accounting for umklapp [circles and dashed lines in Fig.~\ref{fig:2magnons}(b)].

The relation $k_{2m}^{(f)}=2k_m^{(i)}$ seems to suggest that the resonant condition is associated with stimulated magnon emission by the reflected electron, i.e., $k_1=k_2=k_m^{(i)}$. However, because of magnon interactions, bound two-magnon states with complex $k_2=k_1^*$ become formed instead of real $k_1=k_2$~\cite{karbach1998introduction}.
Analysis of the simulation results reveals that the resonantly excited two-magnon states are not such bound states, but rather unbound two-magnon states with $k_1$, $k_2$ shifted 
with respect to $k_m^{(i)}$ in the opposite directions. For instance, for the resonant values $k_e^{(i)}=1.1/a$, $k_m^{(i)}=1.88/a$, we obtain $k_1=1.51/a$, $k_2=2.26/a$ for the reflected electron. This finding provides insight into the problem of stimulated scattering in nonlinear systems, warranting further studies beyond the scope of this work.

\section{IV. Scattering of electron polarized in the x direction}

In this Section, we analyze the scattering of an electron polarized in the $x$ direction, perpendicular to the local spins in their ground state. Absorption of the spin current component perpendicular to the magnetization plays a central role in the semiclassical theories of ST. Naively, this process is unrelated to magnon absorption or emission, which requires transfer of the z-component. Nevertheless, we show that scattering of the x-polarized electron can be interpreted as a superposition of scattering of spin-up and spin-down electrons, as may be expected from the spin decomposition $|s_x\rangle=(|\uparrow\rangle+|\downarrow\rangle)/\sqrt{2}$. As a consequence, the outcomes of this process are determined by the conservation laws discussed above.

\begin{figure}
	\includegraphics[width=\columnwidth]{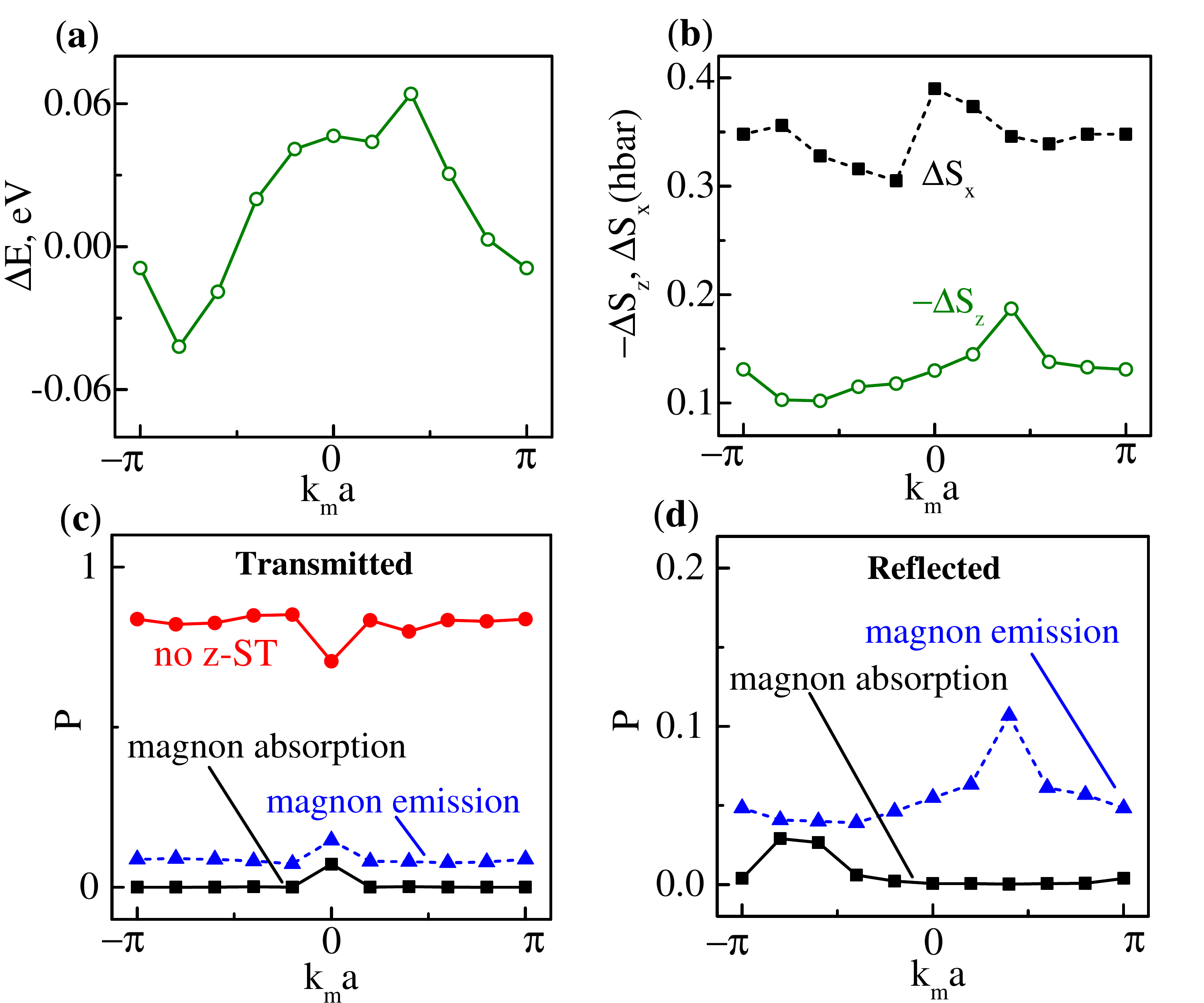}
	\caption{\label{fig:x_polariation} (Color online) Scattering of the electron polarized in the $x$ direction vs $k_m^{(i)}$, for $J=0.2$~eV, $k_ea=0.8$. (a) Energy transfer $\Delta E=E_e^{(i)}-E_e^{(f)}$ from the electron to the local spins. (b) Transfer of the $z$ (open symbols) and $x$ (solid symbols) electron spin components to the local spins. Lines connecting symbols are guides for the eye. (c),(d) Probabilities of different scattering scenarios for the transmitted (c) and reflected (d) electron. Reflection without ST is negligible and not shown.}
\end{figure}

Energy transfer between the electron and the local spins exhibits complex variations with $k_m^{(i)}$, Fig.~\ref{fig:x_polariation}(a). Overall, at small $k_m^{(i)}$ energy flows from the electron to the local spins, while at large $k_m^{(i)}$ it flows from the local spins to the electron. The transfer of the x- and z-components of spin also exhibit a complex dependence on $k_m^{(i)}$, Fig.~\ref{fig:x_polariation}(b). The transferred x-component is always positive, i.e., the initial electron spin is always partially absorbed by the local spins, consistent with the usual approximations of semiclassical ST theories. In contrast, the transferred $z$ component of spin is always negative, indicating that the magnon emission process is dominant over magnon absorption. We now demonstrate that the dependencies in Fig.~\ref{fig:x_polariation}(a),(b) result from the constraints imposed by the conservation laws. 

Figures~\ref{fig:x_polariation}(c),(d) show the probabilities of different scattering scenarios identified in Fig.~\ref{fig:main_schematic}, determined from the quantum simulations by projecting the final state of the system onto the corresponding eigenstates. The magnon absorption probability becomes finite at $k_m^{(i)}=0$ for the transmitted electron component, and at large negative $k_m^{(i)}$ for the reflected electron component, due to the energy and momentum constraints discussed in Section III. The magnon emission probability remains finite for all $k_m^{(i)}$, and exhibits peaks at $k_m^{(i)}=0$ for the transmitted electron component, and at $k_m^{(i)}=1.3/a$ for the reflected component, in accordance with the magnon emission mechanisms discussed in Section III.

The variations of the energy and spin transfer are explained in terms of these contributions, as follows. The minimum in the transferred energy [Fig.~\ref{fig:x_polariation}(a)], coinciding with the maximum of the transferred z-component of spin [Fig.~\ref{fig:x_polariation}(b)], is explained by the resonant absorption of the initial magnon as well as a broad minimum in magnon emission for the reflected electron component [Fig.~\ref{fig:x_polariation}(d)]. On the other hand, a peak in the energy transfer coinciding with a minimum in the transfer of the z-component of spin, is explained by the resonant magnon emission. The negative energy transfer for large initial magnon momentum can be understood as a consequence of the availability of many two-magnon states with energies smaller than that of the initial one-magnon state, which is not the case for small $k_m^{(i)}$ due to the constraints imposed by momentum conservation.

The transfer of the $x$ spin component [solid symbols in Fig.\ref{fig:x_polariation}(b)] does not follow the trends discussed above. To analyze this effect in terms of the conservation laws, the states of the magnetic system must be expanded in the $x$ basis. In the absence of anisotropy or a Zeeman field, a magnon in the $x$ basis is a superposition of two magnons polarized in $z$ and $-z$ directions. Additionally, the ground state of the z-basis is transformed into a highly excited state in the x-basis. Analysis of ST in these highly excited states is beyond the scope of the present work.

\section{V. Conclusions}

In this work, we utilized simulations of electron scattering by a chain of quantum spins, to reveal the quantum constrains on ST that may not be captured by the semiclassical approximation for the magnetic system. Our main results are:

\begin{itemize}\setlength\itemsep{0em}
	\item The generally complex process of ST can be decomposed into a superposition of several distinct processes with qualitatively different characteristics. These processes are magnon absorption, emission, and the magnon number-conserving process, accompanied by electron transmission or reflection,
	\item In addition to angular momentum conservation usually considered in the analysis of ST, energy and momentum conservation laws, in the de Broglie sense of energy and momentum of quasiparticles involved in ST, the electrons and the magnons, substantially constrain the energy and the momentum of magnons that can be absorbed or emitted in the ST process. These constraints may enable the realization of selective cooling of specific magnon modes, or laser-like emission of specific modes not limited to the lowest-frequency dynamical states excited in typical ST experiments,
	\item Even in the absence of ST, spin-polarized electrical current can change the dynamical state of the magnetic system. In particular, we demonstrated the magnon drag effect - the change of the magnon momentum due to the electron current. Generally, ST processes are highly asymmetric with respect to the direction of magnon momentum relative to the electron current,
	\item The magnon emission process, which is central to the ST-induced generation of coherent dynamical magnetization states, involves a complex interaction with the existing magnon in the system, which bears some signatures of the usual stimulated emission of bosons, but is more complex due to the nonlinear magnon interactions, which warrants further theoretical and experimental studies.
\end{itemize} 

The presented analysis was based on numerical simulations of the simplest Heisenberg Hamiltonian and the $s$-$d$ exchange approximation. Because of the exponential scaling of the dimensionality of the Hilbert space with the size of the quantum systems, our simulations were limited to a chain of 10 local spins representing the magnetic system. We expect these results to be directly relevant to quasi-1D systems. However, some of our conclusions will likely become significantly modified in higher dimensions. For instance, in 2D and 3D, the momentum conservation will not limit the possible scattering processes only to reflection or transmission, expanding the range of accessible dynamical states. Interface roughness and defects braking the translational invariance should further relax the constraints imposed by momentum conservation. Nevertheless, based on our analysis, we can conclude that the spatial characteristics of the magnetic systems, such as the quality of their interfaces and the spatial homogeneity, which do not play any role in the semiclassical limit, must significantly affect the efficiency of ST and the spectral characteristics of the dynamical states induced by this effect.

Another demonstrated effect, expected to be relevant to ST regardless of the system dimensionality, is the magnon drag effect - the directionality of the magnon flow induced by ST and  controlled by the direction of electron flow. This effect is highly attractive both for magnonics and for spin-caloritronics. The demonstrated effects are also relevant for the reciprocal phenomena associated with electron transport. For instance, using conservation laws as selection rules, the electron's dynamics can be controlled, enabling almost reflectionless flow of electrons through magnetic systems. Furthermore, certain dynamical magnetization states can serve as electron accelerators, increasing the transmitted electron's velocity due to energy and momentum transfer from the magnetic system. Yet another possibility highlighted by our simulations is electron stopping due to the interaction with magnetic systems, enabling the formation of entangled electron-magnon state that can be useful for the quantum information technologies~\cite{Tejada_2001, Bou_Comas_2019}. More generally, our results suggest that the interaction of conduction electrons with localized moments in magnetic systems may host a plethora of interesting and potentially practically useful, but hitherto largely unexplored phenomena stemming from the quantum nature of magnetism.

This work was supported by the U.S. Department of Energy, Office of Science, Basic Energy Sciences, under Award \# DE-SC0018976.

\bibliography{QST_1M_sim}

\begin{thebibliography}{37}
\bibitem [{com()}]{comment2}%
  \BibitemOpen
  \href@noop {} {}\bibinfo {note} {For ferromagnets, the magnon dsipersion gap
  associated with the magnetic field and anisotropy in real materials is
  negligible on the scale of characteristic electron energies or
  temperature}\BibitemShut {NoStop}%
\end{thebibliography}
\bibliographystyle{apsrev4-1}
\end{document}